\documentclass[11pt]{article}
\pdfoutput=1
\usepackage[export]{adjustbox}
\usepackage{jheppub}
\usepackage[nottoc]{tocbibind}
\usepackage[svgnames]{xcolor}
\usepackage{float}
\usepackage[utf8]{inputenc}
\usepackage{slashed,verbatim}
\pdfoutput=1
\usepackage[most]{tcolorbox}
\usepackage{epigraph,lipsum}
\usepackage{fancyhdr}
\usepackage{epigraph}
\usepackage{graphicx}
\usepackage[normalem]{ulem}

\makeatletter
\def\@fpheader{\relax}
\makeatother

\usepackage{framed}
\usepackage{caption}

\usepackage{lipsum}

\newcommand\blfootnote[1]{%
  \begingroup
  \renewcommand\thefootnote{}\footnote{#1}%
  \addtocounter{footnote}{-1}%
  \endgroup
}
\usepackage{titlesec}
\titleformat*{\section}{\LARGE\bfseries}
\titleformat*{\subsection}{\Large\bfseries}
\titleformat*{\subsubsection}{\large\bfseries}
\titleformat*{\paragraph}{\large\bfseries}
\titleformat*{\subparagraph}{\large\bfseries}

\usepackage[framemethod=tikz]{mdframed}

\usepackage{amsmath,epsfig}
\usepackage{amssymb,amsfonts}
\usepackage{latexsym}
\usepackage{graphicx}

\usepackage{subeqnarray}

\usepackage{graphicx}

\def\be{\begin{equation}}
\def\ee{\end{equation}}
\def\bea{\begin{eqnarray}}
\def\eea{\end{eqnarray}}

\newcommand{\bear}{\begin{eqnarray}}

\newcommand{\eear}{\end{eqnarray}}

\newcommand{\bsea}{\begin{subeqnarray}}
\newcommand{\esea}{\end{subeqnarray}}
\newbox\pippobox

\def\6{\partial}

\newcommand{\comments}[1]{}
\allowdisplaybreaks[3]

\preprint{IFT-UAM/CSIC-19-143}

\begin{document}


\title{\centering \Huge Transverse Collective Modes\\ in Interacting Holographic Plasmas}
\author[\,\dagger]{Matteo Baggioli}
\author[\,\star]{, Ulf Gran}
\author[\,\star]{, Marcus Torns\"{o}}
\vspace{0.1cm}

\affiliation[\dagger]{Instituto de Fisica Teorica UAM/CSIC,
c/ Nicolas Cabrera 13-15, Cantoblanco, 28049 Madrid, Spain}

\affiliation[\star]{Department of Physics, Division for Theoretical Physics, Chalmers University of Technology
SE-412 96 G\"{o}teborg, Sweden}

\emailAdd{matteo.baggioli@uam.es}
\emailAdd{ulf.gran@chalmers.se}
\emailAdd{marcus.tornso@chalmers.se}

\vspace{1cm}

\abstract{We study in detail the transverse collective modes of simple holographic models in presence of electromagnetic Coulomb interactions. We render the Maxwell gauge field dynamical via mixed boundary conditions, corresponding to a double trace deformation in the boundary field theory. We consider three different situations: (i) a holographic plasma with conserved momentum, (ii) a holographic (dirty) plasma with finite momentum relaxation and (iii) a holographic viscoelastic plasma with propagating transverse phonons. We observe two interesting new features induced by the Coulomb interactions: a mode repulsion between the shear mode and the photon mode at finite momentum relaxation, and a propagation-to-diffusion crossover of the transverse collective modes induced by the finite electromagnetic interactions. Finally, at large charge density, our results are in agreement with the transverse collective mode spectrum of a charged Fermi liquid for strong interaction between quasi-particles, but with an important difference: the gapped photon mode is damped even at zero momentum. This property, usually referred to as anomalous attenuation, is produced by the interaction with a quantum critical continuum of states and might be experimentally observable in strongly correlated materials close to quantum criticality, e.g.~in strange metals.
}

\blfootnote{$\dagger$   \,\,\,\url{https://members.ift.uam-csic.es/matteo.baggioli}}

\maketitle

\section{Introduction}
Coulomb interactions play a fundamental r\^ole in the dynamics of condensed matter systems because of their long-range nature. Plasmons, which determine the optical properties of metals and semiconductors, are the most dramatic manifestation of that \cite{nozieres1999theory}. Since the introduction of the plasmon concept by Pines and Bohm \cite{1952PhRv...85..338P} as quantized bulk plasma oscillations in metallic solids, a lot has been done and several potential technological applications have been devised, \color{black} e.g.~optical devices for information technology, sensing, nonlinear optics, optical nanotweezers, biomedical applications and renewable energy technologies \cite{maier2007plasmonics}\color{black}.

In principle, the entirety of plasmon physics can be described and understood using Maxwell's equations in a medium and introducing the idea of a dielectric function $\epsilon(\omega,k)$ \cite{jackson1975classical}, where $\omega$ is the frequency and $k$ is the wave-vector. However, recent experiments on unconventional, or ``strange'', metals have posed new challenges to this theoretical framework \cite{Mitrano5392,husain2019crossover}. In more specific terms, the quasi-particle nature of plasmons, and the idea of the Lindhard continuum, are not manifest in these cases.
These new exciting experimental results stimulated a lot of activity in the holographic community \cite{Aronsson:2017dgf,Aronsson:2018yhi,Gran:2018vdn,Gran:2018jnt,Mauri:2018pzq,Krikun:2018agd,Romero-Bermudez:2019lzz,Baggioli:2019aqf,Andrade:2019bky}, driven by the hope of describing these new phenomena within the framework of strongly coupled theories and their gravitational duals. A first positive outcome of this program has been the observation of an anomalous attenuation, or damping, of the plasmon mode at $k=0$, in contrast to the result from Fermi-liquid theory where the mode is protected from decay by the absence of any available decay channel. This was first observed in \cite{Aronsson:2017dgf}, and later elaborated on in \cite{Krikun:2018agd}. The residual damping at $k=0$ is natural in holographic models due to the absence of quasi-particles, and the related appearance of an incoherent critical continuum of excitations. This leads to the experimentally observed featureless spectrum in strange metals \cite{Mitrano5392,husain2019crossover}, where a highly damped plasmon is only visible at low momentum. Furthermore, the possibility of having new dispersion relations, especially linked to the $k$-gap phenomenon \cite{2016RPPh...79a6502T,2017PhRvL.118u5502Y,Baggioli:2018vfc,Baggioli:2018nnp,Baggioli:2019jcm,Davison:2014lua}, has been discussed \cite{Gran:2018vdn,Baggioli:2019aqf}.

Inspired by these questions, in this manuscript, we discuss the dynamics of the transverse spectrum of charged holographic systems in presence of long-range Coulomb interactions. It is a well-known fact \cite{nozieres1999theory} that the transverse collective modes of a charged system are coupled to the electromagnetic waves. The dispersion relation for the coupled ``electromagnetic collective
modes'' can be obtained directly once we know the transverse
conductivity, using the following expression
\begin{equation}
c^2\,k^2\,=\,\omega^2\,\epsilon_{\perp}(\omega,k)\,=\,\omega^2\,+\,4\,\pi\,i\,\omega\,\sigma_{\perp}(\omega,k)\,,\label{full}
\end{equation}
where $c$ is the speed of light.  \color{black} The transverse conductivity $\sigma_\perp(\omega,k)$ is defined in terms of the two-point function of the transverse current
\begin{equation}
    \sigma_\perp(\omega,k)\,\equiv\,\langle J_y\,J_y \rangle (\omega,k)
\end{equation}
upon assuming the momentum being aligned in the $x$ direction. In this notation, $\sigma_\perp(\omega,0)$ is the standard electric conductivity. \color{black}
It's ``only'' a matter of solving one exercise proposed in the book of Pines and Nozi\`eres \cite{nozieres1999theory} to verify that, in Fermi liquids, the transverse collective
mode only exists for strong repulsion between quasi-particles,
namely when the first Landau parameter $F_1^s > 6$. \color{black}From a physical perspective, the strong coupling, which corresponds to a large renormalized mass, is necessary to separate the propagating transverse collective mode from the particle-hole continuum living below $\omega= v_F k$, with $v_F$ the Fermi velocity \cite{PhysRevB.99.075434}.  This situation can appear in systems close to criticality where the effective mass $m^*$ diverges. Even more interestingly, this might be the case in the context of strongly correlated metals such as those appearing in the experiments of \cite{Mitrano5392,husain2019crossover}, where weakly coupled Fermi liquid theory is likely not applicable. \color{black}
\begin{figure}[h]
\center
\includegraphics[width=0.4\linewidth]{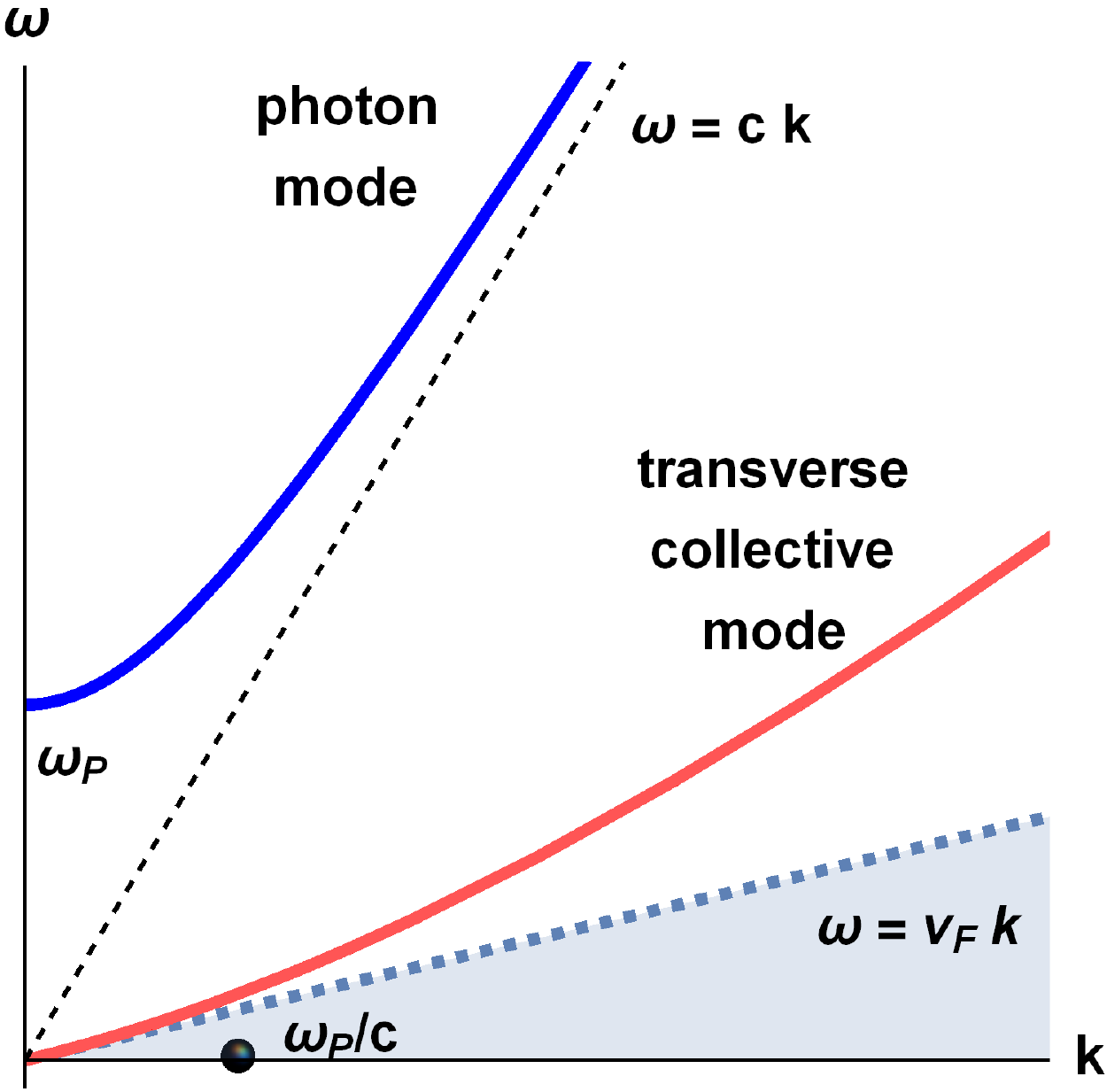}
\caption{The dynamics of the transverse modes. The shaded region represents the continuum of single-pair excitations. The black dot indicates the point at which the transverse collective mode disappear into the continuum. The Fermi velocity is indicated with $v_F$. Figure adapted from \cite{nozieres1999theory}.}
\label{fig:Pines}
\end{figure}

Before proceeding to the model and the results, let us briefly summarize what is expected from a quasi-particle point of view. A transverse deformation in a material produces a polarization, which itself interacts with the electromagnetic field. The EM response provides a feedback mechanism in the sample and together this gives rise to the transverse collective modes. The full spectrum of modes can be obtained by solving \eqref{full}; here, we comment only on the most salient features shown in figure~\ref{fig:Pines}. \color{black} First, let us consider the large frequency regime $\omega \gg v_F \,k$, away from the particle-hole continuum which in figure~\ref{fig:Pines} is indicated by the shaded region. There, the dielectric function can be well approximated by the simple form
\begin{equation}
    \epsilon_\perp(\omega,k)\,=\,1\,-\,\frac{\omega_P^2}{\omega^2} \label{die}
\end{equation}
where $\omega_P$ is the plasma frequency, which is microscopically given by
\begin{equation}
    \omega_P^2\,\equiv\,\frac{n_e\,e^2}{\epsilon_0\,m^*}~,
\end{equation}
with $n_e$ the electron density, $e$ the EM coupling, $\epsilon_0$ the dielectric constant of the vacuum and $m^*$ the effective mass. Plugging \eqref{die} into the master equation \eqref{full}, the solution becomes
\begin{equation}
    \omega^2=\omega_P^2+c^2\,k^2~,
\end{equation}
which is the standard dispersion relation for a plasma oscillation with a mass gap given by the plasma frequency $\omega_P$.
At large momentum compared to the plasma frequency, i.e.~$c\, k \gg \omega_P$, the coupling between the EM waves and the electrons is very weak and the dispersion relation of the modes is not modified. In that regime, we have a \textit{photon root} ~$\omega= c \,k$ and a propagating ''electronic shear sound mode''~$\omega= v_s \,k$, where the speed of sound is given by the elastic properties of the electronic liquid, namely the Landau parameters \cite{1999PhRvB..60.7966C}. The major effects of the EM interactions occur in the so-called strongly coupled region at low momentum $c \,k \ll \omega_P$. In that regime, the ''photon root'' becomes gapped. The same is true for the transverse collective mode which at low momentum is repelled by the photon root and disappears in the continuum, as shown in figure~\ref{fig:Pines} for $k < \omega_P/c$.\color{black}

\color{black}Interestingly, the transverse collective mode should be present in a large class of interacting charged and neutral Fermi liquids especially those in proximity to critical points where the quasiparticle mass diverges \cite{PhysRevB.99.075434}. Recently, a viable experimental setup for its detection has been proposed \cite{khoo2020quantum}. According to the results of that paper, ``\textit{the shear
sound is responsible for the appearance of sharp dips in the AC conductance of narrow channels at resonant frequencies matching its dispersion. Ultra-clean 2D materials that can be tuned towards the Wigner crystallization transition such as Silicon MOSFETs, MgZnO/ZnO, p-GaAs and AlAs quantum wells are promising platforms to experimentally discover the shear sound.}''\color{black}

Finally,
transverse collective modes have been discussed in dusty (dirty) plasmas and Yukawa fluids both in simulations \cite{PhysRevLett.84.6026,doi:10.1063/1.5088141} and experiments \cite{PhysRevLett.97.115001}. Interestingly enough, the appearance of a $k$-gap in the dispersion relation of the transverse modes has been identified, in good agreement with the holographic results.

\section{The model}\label{sec:model}
We consider the holographic bottom-up model introduced in \cite{Baggioli:2014roa,Alberte:2015isw} and described by the following action
\begin{equation}\label{eq:action}
S\,=\,\int d^4x \sqrt{-g}
\left[\frac{R}2+\frac{3}{\ell^2}- \,\frac{1}{4}\,m^2 \,V(X)\,-\,\frac{1}{4}\,F^2\right]~.
\end{equation}
The model consists of the standard Einstein-Maxwell terms plus an additional scalar sector defined in terms of an arbitrary potential of the quantity $
    X\,\equiv\,\partial_\mu \phi^I \partial^ \mu \phi^I$, where $\phi^I= \color{black}\alpha\color{black}x^I,\,\text{with}\, I=x,y$. This system admits a simple asymptotic AdS black-brane solution with metric
\begin{equation}
\label{backg}
ds^2=\frac{1}{z^2} \,\left[\,-f(z)\,dt^2\,+\,\frac{dz^2}{f(z)}\,+\,dx^2\,+\,dy^2\,\right]~,
\end{equation}
with the holographic $z$ coordinate spanning between the AdS boundary $z=0$ and the black hole horizon $z=z_h$ (defined as $f(z_h)=0$). The bulk profiles of the gauge and scalar fields \color{black} have simple solutions,
\begin{align}
   & A_t(z)\,=\,Q\,(1-\,z)\,,
\qquad \phi^I\,=\,\alpha\,x^I\, ,
\end{align}
where $Q$ and $\alpha$ parameterize the background, and the emblackening factor $f$ is defined as
\begin{equation}\label{backf}
f(z)= z^3 \int_z^{z_h} dv\;\left[ \frac{3}{v^4} -\frac{m^2}{4\,v^4}\, 
V(v^2)\,-\,\frac{Q^2}{2 z_h^4} \right] \, .
\end{equation}
With the holographic dictionary, the chemical potential $\mu$ and the charge density $\rho$ of the dual field theory can be read off as
\begin{equation}
    \mu			\;=\;	\frac{Q}{\sqrt{\lambda}\, z_h}	\, , \qquad
\rho	\;=\;	\frac{\sqrt{\lambda}\, Q}{z_h^2}	\, ,\label{eq:muandrho}
\end{equation}
where we have introduced $\lambda$, a relative coupling strength for the electromagnetic field between the bulk and boundary theories as in \cite{Aronsson:2017dgf}, 
\begin{equation}
    F_{boundary}=\frac{1}{\sqrt{\lambda}}F_{bulk}|_{\partial M}\,,\quad J_{boundary}=\sqrt{\lambda}\,\partial_{z}A_{bulk}|_{\partial M}~,
\end{equation} 
i.e.~increasing $\lambda$ means a smaller EM-field on the boundary and a larger charge. \color{black}
Finally, the temperature of the dual field theory is given by
\begin{equation}
T=-\frac{f'(z_h)}{4\pi}=\frac{12 -  m^2\, V\left(z_h^2 \right)\,-\,2\,Q^2 }{16\, \pi \,z_h}\,.\label{eq:temperature}
\end{equation}

In this manuscript, we consider the transverse sector of the fluctuations (see details in appendix \ref{app1}). Importantly, following the prescription given in \cite{Aronsson:2018yhi}, we get the mixed boundary conditions
\begin{equation} 
   \left( \omega^2\,-k^2\right)\delta\!A_y^{(0)}\,+\,\lambda\,\delta\!A_y^{(1)}\,=\,0\,\label{eq:bcs}
\end{equation}
for the gauge field perturbation, whose asymptotic behaviour turns out to be
\begin{equation}
    \delta\!A_\mu\,=\,\delta\!A_\mu^{(0)}\,+\,\delta\!A_\mu^{(1)}\,z\,+\,\mathcal{O}(z^2)~.
\end{equation}
The mixed boundary conditions in \eqref{eq:bcs}, which are equivalent to a double-trace deformation \cite{Aronsson:2017dgf,Mauri:2018pzq}, make the gauge field dynamical \color{black} in the boundary field theory\footnote{\color{black}Using the holographic dictionary, it follows that, under standard boundary conditions, considering a local $U(1)$ gauge symmetry in the bulk corresponds to having a global $U(1)$ symmetry in the dual field theory. In this sense, there are no EM interactions in the boundary theory since there is no dynamical photon.\color{black}} \color{black}and implement the effects of the electromagnetic Coulomb interaction. In this sense, the parameter $\lambda$ can indeed be associated with the strength of Coulomb interactions. The speed of light $c$ in \eqref{full} has here been set to unity (as a choice of convention).

In the following, we will keep $\lambda$ finite (unless otherwise mentioned we set $\lambda=1$) and we will consider three different situations
\begin{itemize}
    \item[(I)]: $m^2=0$. This is the simple Einstein-Maxwell model. The dual field theory represents a charged relativistic plasma with Coulomb interactions.
    \item[(II)]: $m^2 \neq 0$ and $V(X)=X$. This is the famous \textit{linear axion model} \cite{Andrade:2013gsa}. The dual field theory represents a charged relativistic plasma with Coulomb interactions and momentum relaxation. The momentum dissipation rate will be determined by the value of the dimensionless quantity $\alpha/T$.
    \item[(III)]: $m^2 \neq 0$ and $V(X)=X^3$. This model breaks translations spontaneously and displays the presence of transverse (and longitudinal) propagating phonons \cite{Alberte:2017oqx,Ammon:2019apj}. The dual field theory represents a charged relativistic plasma with Coulomb interactions and finite elastic moduli. The rigidity of the system is parameterized by the dimensionless quantity $\alpha^3/T$.
\end{itemize}
\section{Transverse collective modes}
\subsection{A relativistic charged plasma with Coulomb interactions}
\begin{figure}[h]
    \centering
    \includegraphics[width=0.49 \linewidth]{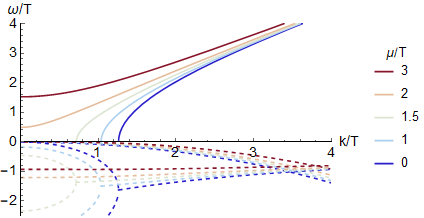}
      \quad   \includegraphics[width=0.45 \linewidth]{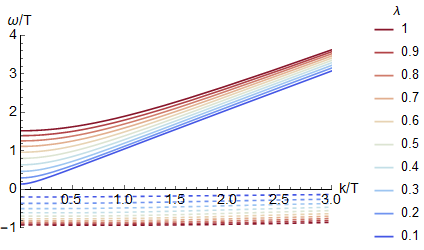}
     
    \caption[]{The transverse collective modes in absence of the scalar sector. \color{black} Solid lines are the real parts of the modes\footnotemark, and dashed lines the imaginary parts. \textbf{Left: } The modes at different values of the dimensionless chemical potential $\mu/T$. Note especially that the imaginary part (here mostly horizontal) of the mode with non-zero real part breaks apart into two purely imaginary modes at small $k$ and $\mu$. \color{black} \textbf{Right: } The mode in the Maxwell sector at $\mu/T=3$ as a function of the EM coupling $\lambda$.}
    \label{fig:noaxions}
\end{figure}

We start by considering a relativistic charged plasma with finite EM interactions. The gravitational dual picture is a charged Reissner-Nordström (RN) black hole with modified boundary conditions as in \eqref{eq:bcs}, i.e.~at finite $\lambda$, for the gauge field perturbations. The numerical results \color{black} for the two modes with lowest energy \color{black} are shown in figure~\ref{fig:noaxions}.

At zero charge, $\mu=0$, the gravitational and the Maxwell sectors are decoupled. The dynamics of the gravitational sector is simply that of a shear diffusive mode $\omega=-i \,D\,k^2$ \cite{Policastro:2002se}, where, as usual, the diffusion constant is set by the universal value of the shear viscosity $D=\eta/sT= 1/4\pi T$ \cite{Kovtun:2012rj,Policastro:2001yc}. The dynamics of the Maxwell sector displays a typical $k$-gap dispersion relation (see~\cite{Baggioli:2019jcm} for a recent review) which can be understood directly from the master equation \eqref{full}. Because of the absence of Galilean invariance, the resistivity is non-zero even at zero charge density, where it is governed by the so-called incoherent conductivity $\sigma(\omega)=\sigma_0$ \cite{Davison:2015taa}\footnote{\color{black} Galilean invariance constraints the momentum density $T^{ti}$ to be proportional to the electric current $J^i$. This requires that $\sigma_0=0$. Physically, a non-zero $\sigma_0$ corresponds to the possibility of creating electron-hole pairs moving in different directions. The pairs would not transport momentum, but they would carry charge and therefore contribute to the conductivity. This also explains why Galilean
invariance is broken in our model as the number of electrons is not a conserved quantity,
due to the creation/annihilation of the electron-hole pairs.\color{black}}. Inserting a constant value for the conductivity in formula \eqref{full}, one recovers immediately the $k$-gap behaviour shown in figure~\ref{fig:noaxions}. To the best of our knowledge this feature first appeared in \cite{Arias:2014msa}, and was then re-discovered in the context of global symmetries in \cite{Hofman:2017vwr}.
\footnotetext{\color{black} It should be noted that due to symmetry, there are also modes with negative real parts, but as they are symmetric around the $k/T$-axis, they are omitted to reduce clutter in the graph.\color{black}}

After introducing a finite chemical potential, the two sectors are not decoupled anymore and they start interact with a strength proportional to the charge density itself. The effect on the shear mode is not dramatic and simply leads to a rescaling of the diffusion constant $D$ \cite{Ge:2008ak}. The dynamics in the Maxwell sector is more interesting. A finite charge density leads to several consequences for the $k$-gap dynamics: Firstly, the hydrodynamic charge diffusion mode is given a finite damping at $k=0$, and therefore ceases to be a hydrodynamic mode in the strict sense. The situation is analogous to what happens in the spectrum of the linear axion model \cite{Andrade:2013gsa} once translations are broken \cite{Baggioli:2018vfc,Baggioli:2018nnp}. Secondly, the position of the momentum gap moves towards lower momenta, until approaching the origin and being converted to a real frequency/energy gap (a mass) (see \cite{2019NatSR...9.6766T} for a field theory analysis). Thirdly, the separation between the two colliding modes decreases with increasing charge density and it eventually shrinks to zero at the point where the momentum gap closes. At that specific point, one recovers a linear and massless propagating photon mode, $\text{Re}[\omega] = c\,k$. Increasing the chemical potential further, the charge density produces a gapped mode whose mass is determined by the plasma frequency\footnote{To be precise, there is also a contribution to the mass gap from the \color{black} momentum relaxation rate $\Gamma$\color{black}, which becomes subdominant in the limit of large charge density.} $\omega_P$. At this point, the spectrum displays a typical quadratic dispersion $\omega^2=\omega_P^2\,+\,c^2\,k^2$, since the efficient screening from the large charge density results in weaker long-range correlations, and therefore physics resembling that of conventional Fermi-Landau theory. \color{black} Increasing the relative coupling $\lambda$, which corresponds to increasing the charge density, makes the gap larger (right panel of figure~\ref{fig:noaxions}). \color{black} Note, however, that this mode is strongly damped, even at zero momentum $k=0$. This implies the presence of an extra damping mechanism in addition to the common Landau damping, as already discussed for the longitudinal case in \cite{Aronsson:2017dgf,Krikun:2018agd}, which is sometimes called an \textit{anomalous attenuation}. Its origin lies in the presence of a quantum critical continuum of states, which holographically can be understood by the mode originating from the collision of two highly damped non-hydrodynamical modes \cite{Gran:2018vdn}, and has potentially been experimentally observed in \cite{Mitrano5392,husain2019crossover}.

\color{black}Let us spend a few words on the diffusion-to-sound crossover, usually referred to as the $k-$gap phenomenon \cite{Baggioli:2019jcm}. The appearance of a relaxation time responsible for this effect can be understood directly from the master equation \eqref{full} as it can be rewritten as
\begin{equation}
    \omega^2\,+\,i\,\frac{\omega}{\tau}\,=\,c^2\,k^2\,+\,\dots
\end{equation}
which gives rise to the aforementioned collision at $k=1/(2 c \tau)$,
\begin{equation}
    \omega_\pm\,=\,-\,\frac{i}{2\,\tau}\,\pm\,\sqrt{c^2\,k^2\,-\,\frac{1}{4\,\tau^2}}\,.
\end{equation}
Writing the master equation as
\begin{equation}
    \epsilon(\omega,k)\,\omega^2\,=\,c^2\,k^2\,,\quad  \epsilon(\omega,k)\,\equiv\,1\,+\,4\,\pi\,i\,\sigma_\perp(\omega,k)
\end{equation}
it is clear that the existence of a finite relaxation time is due to the polarization of the material, as encoded in a non-trivial dielectric function. It is tempting to associate this relaxation time with how long it takes for the material to lose its polarization -- the dielectric relaxation time. \color{black}

\subsection{A dirty plasma with Coulomb interactions}
In this second case, we add momentum relaxation to our initial plasma by coupling it to a scalar sector. Momentum relaxation will make the plasma ``dirty'' and closer to the realistic situation in dusty plasma, as in \cite{PhysRevLett.84.6026,doi:10.1063/1.5088141,PhysRevLett.97.115001}.

The relaxation of momentum produces a finite damping for the shear mode, whose dispersion relation now becomes
\begin{equation}
    \omega\,=\,-\,i\,\Gamma\,-i\,D\,k^2\,+\,\dots\label{eq:Gamma}
\end{equation}
where $\Gamma$ is the momentum relaxation rate, governed in our case by the dimensionless factor $\alpha/T$, which is proportional to the mass of the graviton. In absence of interactions with the EM sector, \color{black} increasing the momentum relaxation rate $\sim \alpha/T$\color{black}, this mode will move down the imaginary axis, becoming overdamped and extremely short-lived.

When turning on a finite, but small, charge density something interesting happens. At zero momentum relaxation, i.e.~$\alpha/T$=0, and finite but small charge density, we have the following set of modes:
\begin{align}
    &\text{shear mode (gravitational sector)}\quad \rightarrow \quad \omega\,=\,-\,i\,D\,k^2 \quad :\quad \includegraphics[height=1.8cm, valign=c]{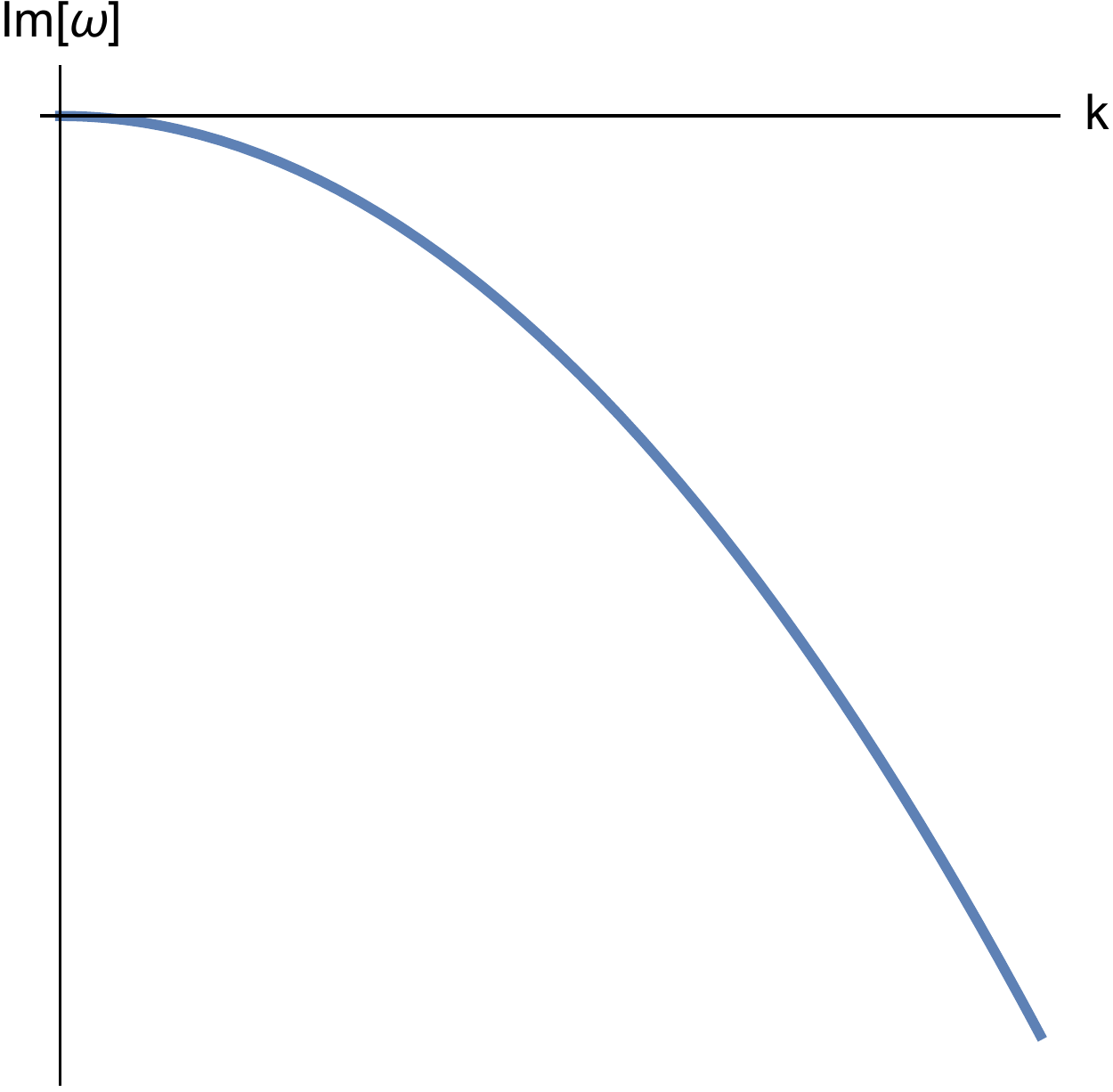}  \label{un}\\
    &\nonumber\\
    &\text{photon mode (Maxwell sector)}\quad \rightarrow \quad \text{damped $k$-gap}\quad :\quad \includegraphics[height=1.8cm, valign=c]{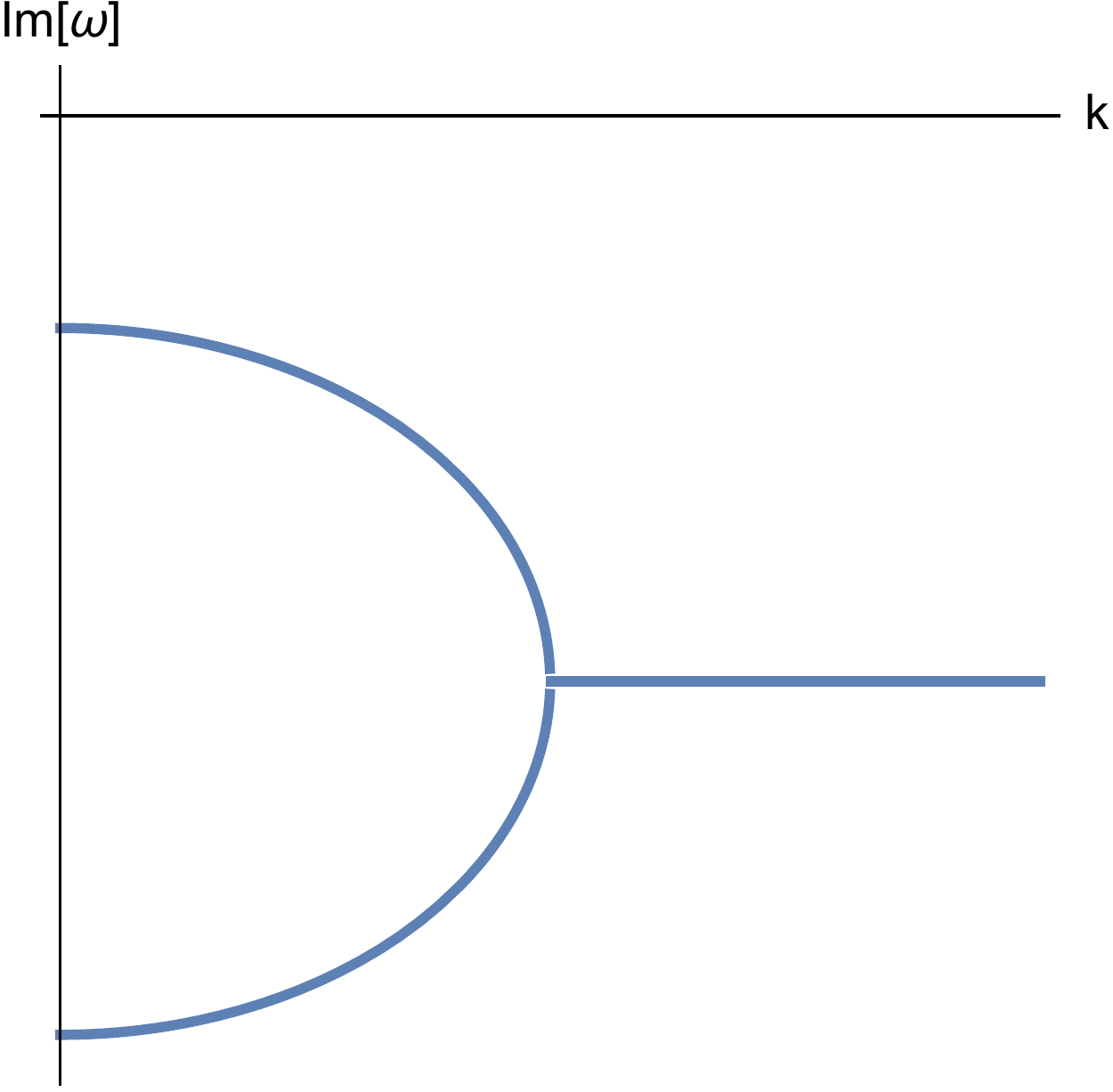} \label{du}
\end{align}

\begin{figure}[h]
    \centering
    \includegraphics[width=0.32 \linewidth]{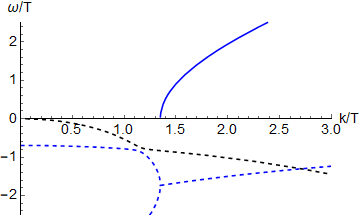}
    \includegraphics[width=0.32 \linewidth]{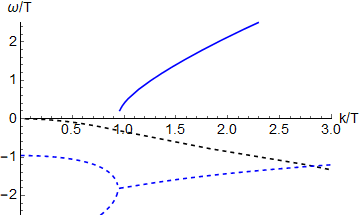}
     \includegraphics[width=0.32 \linewidth]{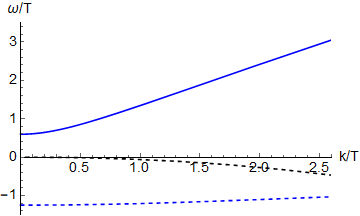}
       \caption{\color{black} Dispersion relations of collective modes at fixed momentum relaxation rate $\Gamma$, encoded in the dimensionless parameter $\alpha/T=3$ (more precisely $\Gamma= \alpha^2/2\pi T$). Solid lines are the real parts, dashed lines imaginary. Similarly colored dashed and solid lines correspond to the same mode. \color{black} \textbf{Left panel: } For small chemical potential, $\mu/T=0.1$. \textbf{Central panel:} For intermediate chemical potential, $\mu/T=1$. \textbf{Right Panel: }For large chemical potential, $\mu/T=2$.}
    \label{fig:dirty}
\end{figure}

\begin{figure}[b]
    \centering
    \includegraphics[width=0.48\linewidth]{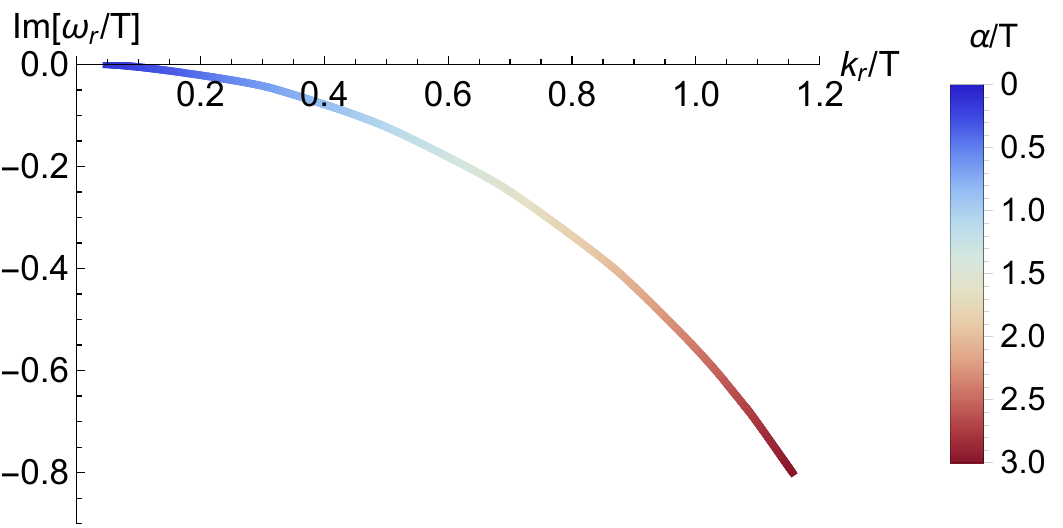}
    \quad
    \includegraphics[width=0.48\linewidth]{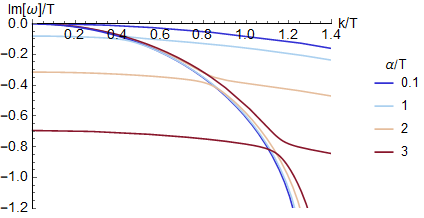}
    \caption{The mode repulsion at different $\alpha/T$. \textbf{Left:} The repulsion point $(k_r,\omega_r)$ for different $\alpha/T$ at $\mu/T\ll 1$. Similar behaviour appears for different $\mu/T$. \textbf{Right:} Dispersion relations highlighting gap opening at $\mu/T=0.1$ near some repulsion points of the left panel.}
    \label{fig:crossing}
\end{figure}

Increasing momentum relaxation, the mode in \eqref{un} moves down and approaches the top mode in \eqref{du}. When they get close to each other, they mix, and then they repel each other as shown in the left panel of figure~\ref{fig:dirty}. This avoided crossing mechanism is quite natural in interacting condensed matter systems (see for example the dispersion relation of polaritons \cite{PhysRev.33.195,1951Natur.167..779H}), but to the best of our knowledge it has never been mentioned in the context of transverse collective modes before. 
\color{black} 
At larger charge density, the EM branch shown in \eqref{du} moves down the imaginary plane and it has less chance to interact with the shear mode (see central panel of figure~\ref{fig:dirty}). The repulsion mechanism which we are discussing happens only in the regime of parameters where the two modes are close to each other and potentially crossing. 
At very large charge density, compared to temperature and to the momentum relaxing parameter $\alpha$, the system does behave as if momentum was conserved. In simple terms, the strong repulsion between the modes pushes the pseudo-diffusive shear mode towards the origin making its damping term vanish. A gapped mode appears, as in the right panel of figure~\ref{fig:dirty}, which has also a finite damping because of momentum relaxation.
At exactly zero chemical potential, the equations of motion decouple, and there is a crossing rather than a repulsion, and a well defined point where the modes cross. These points are shown as a curve in the left panel of figure~\ref{fig:crossing}, and the gaps produced at some specific values of the momentum relaxation and small chemical potential are shown in the right panel. We find that the position of the repulsion point as a function of the same parameter $\alpha/T$ moves farther away from the origin (larger $\omega/T$ and $k/T$) with increasing momentum relaxation. This is quite expected as, to zeroth order, the position of the repulsion point is proportional to the gap $\Gamma$ of \eqref{eq:Gamma}. Additionally, we find that when $\alpha\gg\mu\neq0$, the size of the gap scales with $\alpha$, and for $\alpha\ll\mu$, the gap becomes the same as indicated in \eqref{du}, near $k=0$. 

\begin{figure}[h!]
    \centering
     \includegraphics[width=0.48\linewidth]{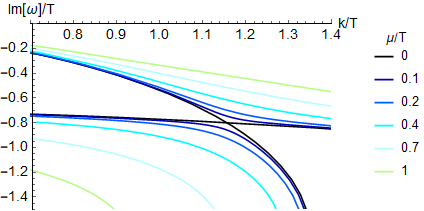}
     \quad
     \includegraphics[width=0.48 \linewidth]{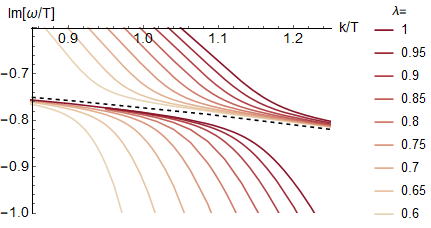}
    \caption{\color{black} Zoom on the dispersion relations at $\alpha/T=3$. \textbf{Left:} The opening of the gap at different values of $\mu/T$ at fixed $\lambda=1$. \textbf{Right:} The gap for different strengths of EM coupling $\lambda$, at $\mu/T=0.1$. The dashed line indicates the decoupled mode found at $\mu/T=0$ (independent of $\lambda$).\color{black}}
    \label{fig:chempot}
\end{figure}

The dependence on the EM interactions, determined by $\mu/T$ and $\lambda$, is shown in  figure~\ref{fig:chempot}.
In the left panel we show the dynamics of the repulsion as a function of $\mu/T$. By increasing the chemical potential of the system, the repulsion between the two modes becomes stronger and the damped shear mode is pushed towards the origin. At the same time, the other modes acquires a bigger imaginary part and the gap between the two becomes larger. Instead, tuning $\lambda$ and keeping fixed $\mu/T$, the size of the gap is kept fairly constant, but moves the position of the gap horizontally, as shown in the right panel of figure~\ref{fig:chempot}. Note however, that this analysis is made specifically at constant $\mu/T$ and varying $\rho/T$. The dictionary takes into account the EM coupling $\lambda$, see \eqref{eq:muandrho}. This means that the effect on the gap from increasing $\lambda$ is primarily the resulting effect from the change in chemical potential, and not from the change in charge density. This analysis confirms that the repulsion mechanism between the shear mode and the charge one is driven by the EM interactions, specifically, the chemical potential. 

\color{black}

\subsection{An elastic plasma with Coulomb interactions}
Let us now consider a different situation, namely a plasma where translations are not broken explicitly but instead spontaneously. This case can be considered by changing the potential in the scalar sector \cite{Alberte:2017oqx}. The appearance of transverse phonon modes is guaranteed by the SSB pattern and has been checked and studied in several works \cite{Alberte:2017cch,Andrade:2019zey,Ammon:2019wci,Baggioli:2019aqf,Ammon:2019apj,Baggioli:2019abx,Baggioli:2019mck,Baggioli:2019elg}. Here, we are interested in analyzing the interaction between the transverse collective sound mode and the transverse EM waves in presence of Coulomb interactions. As explained in the introduction, it is clear that the phonons and photons will interact, leading to the existence of non-trivial collective modes.

\begin{figure}[h!]
    \centering
    \includegraphics[width=0.32 \linewidth]{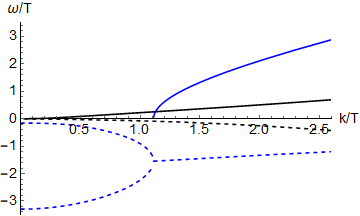}
     \includegraphics[width=0.32 \linewidth]{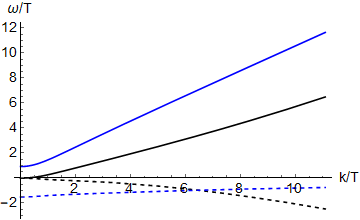}
     \includegraphics[width=0.32 \linewidth]{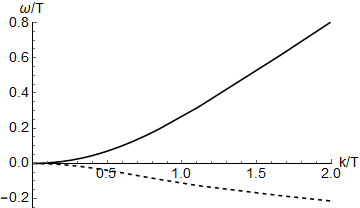}
     \caption{The transverse collective modes in presence of transverse phonons. Dispersion relations as functions of the shear elastic modulus of the system $G$, encoded in the dimensionless parameter $\alpha^3/T$. \color{black} Solid lines are the real parts, dashed lines the imaginary parts. Similarly colored dashed and solid lines correspond to the same mode. \textbf{Left: }Intermediate chemical potential, $\mu/T=1$, and $\alpha^3/T=1$. \textbf{Center: }Large chemical potential, $\mu/T=3$, and $\alpha^3/T=3$.  \textbf{Right: } Zoom on the transverse phonon mode in the center panel.}
    \label{fig:phonons}
\end{figure}

At small chemical potential, the mixing between the two sectors is not very strong, and the modes are rather unaffected by the presence of each other (see left panel of figure~\ref{fig:phonons}). The situation is different in presence of a large chemical potential as shown in the \color{black} center \color{black} panel of figure~\ref{fig:phonons} (additionally, increasing the interaction strength $\lambda$ has a similar effect, producing a similar behaviour to figure \ref{fig:noaxions}). The photon mode becomes massive, with a finite plasma frequency $\omega_P$ related to the charge density and the EM interactions strength $\lambda$. The transverse sound becomes strongly modified at small momentum, $k/T<1$, because of the interaction with the photon mode, \color{black} and instead displays a quadratic behaviour, highlighted in the right panel of figure~\ref{fig:phonons}. The photon mode repels the sound mode as explained in the introduction. In contrast, at large momentum $k/T \gg 1$, the interaction between the two modes is very weak and the modes follow their natural dispersion relations
\begin{equation}
    \omega\,=\,c\,k\,,\quad \omega\,=\,v_{sound}\,k \,,
\end{equation}
where $v_{sound}$ is determined by the rigidity of the system and therefore depends on the dimensionless parameter $\alpha^3/T$.

The larger the parameter $\alpha^3/T$, the larger the shear elastic modulus (computed as introduced in \cite{Alberte:2015isw,Alberte:2016xja}) and therefore also the speed of transverse phonons. The real part of the dispersion relation of the modes is in agreement with the expectations from Fermi liquid theory at strong coupling, shown in figure~\ref{fig:Pines}, but there is a crucial difference in the imaginary part. There is an anomalous damping mechanism which is active even at zero momentum $k=0$ (where the standard Landau damping is ``frozen''). This is the analogue of the \textit{anomalous attenuation} already discussed for longitudinal plasmons in \cite{Aronsson:2017dgf,Krikun:2018agd}, and potentially observed in strange metal experiments in \cite{Mitrano5392,husain2019crossover}.

\begin{figure}[h!]
    \centering
    \includegraphics[width=0.29 \linewidth]{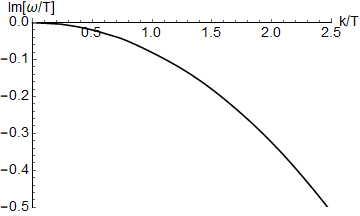} \quad 
    \includegraphics[width=0.29 \linewidth]{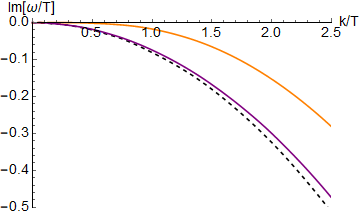} \quad
    \includegraphics[width=0.32 \linewidth]{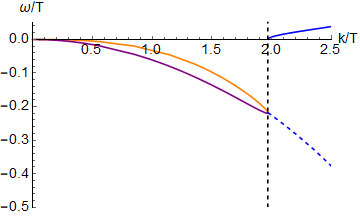}
      \caption{The dynamics of the hydrodynamic modes in the case of an elastic plasma, with the diffusive charge mode omitted. \textbf{Left :} At zero chemical potential and zero elasticity, the diffusion modes of the transverse collective excitations superimpose each other. \textbf{Center :} At finite chemical potential and small elasticity, the two diffusion constants become different and smaller than the $\mu=0$ value (dashed line). \textbf{Right :} Increasing the rigidity of the system, the two modes collide and produce a propagating sound mode in a $k$-gap fashion. The higher the rigidity, parameterized by $\alpha^3/T$, the more the momentum gap moves towards the origin and eventually disappears. \color{black}In this sense, the physics in the central and right panels is qualitative the same. See figure~\ref{fig:merging} for a detailed analysis of the recombination point as a function of $\alpha^3/T$. \color{black}}
    \label{fig:splitting}
\end{figure}

In this last case, i.e.~in presence of propagating transverse modes, we observe another interesting feature which reflects the non-trivial effects of the EM interactions.
At finite charge density (or equivalently chemical potential), the interactions, whose strength are controlled by the parameter $\lambda$, induce a propagation-to-diffusion crossover and the transverse collective modes become diffusive at low momenta. The diffusion constants of the two modes are not equal (\color{black}central panel \color{black}of figure~\ref{fig:splitting}) but exhibit a small splitting. At large enough momentum, these two diffusive modes collide in a $k$-gap fashion and they produce the propagating transverse sound mode, with speed determined by the rigidity of the system (right panel of figure~\ref{fig:splitting}).

\begin{figure}[h!]
    \centering
    \includegraphics[width=0.49 \linewidth]{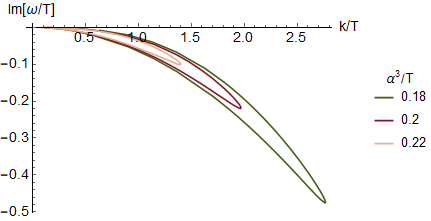}
    \includegraphics[width=0.49 \linewidth]{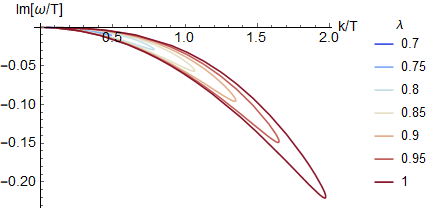}
      \caption{The pole collision at $\mu/T=3$. \textbf{Left:}  Zoom on the dispersion relations before the mode collision for different values of the elasticity $\alpha^3/T$. \textbf{Right:} Zoom on the dispersion relations before the mode collision for different values of the EM coupling $\lambda$, at $\alpha/T=0.2$.}
    \label{fig:merging}
\end{figure}

The dynamics is shown in figure~\ref{fig:splitting}. At zero chemical potential, and zero elasticity $\alpha^3/T=0$, the two diffusion modes superimpose with the same diffusion constant $D=1/(4\pi T)$ (see the left panel of figure~\ref{fig:splitting}). Increasing the chemical potential, the diffusion constants of the transverse collective modes become smaller and take on different values. Increasing the rigidity of the system, parameterized by $\alpha^3/T$, the two diffusive modes collide at a certain critical momentum (see panel c of figure~\ref{fig:splitting}). At momenta larger than the critical value, a propagating transverse phonon with $Re(\omega) \neq 0$ appears. Moreover, we find that the $k$-gap becomes smaller by increasing the rigidity of the system. The details of this mechanism are shown in figure~\ref{fig:merging}. Importantly, we observe that this collision is a result of the Coulomb interaction. As shown in the right panel of figure~\ref{fig:merging}, this feature disappears in absence of Coulomb interactions, $\lambda \rightarrow 0$, and it becomes more and more pronounced at large EM coupling $\lambda \sim 1$. Therefore, the effects of the EM interactions is to inhibit the propagation of the transverse collective modes at large distances, which is natural given the dissipative dynamics present even at $k=0$ as a result of the absence of quasi-particles, and hence an incoherent continuum of available excitations. 

\begin{figure}[h]
\center
\includegraphics[width=0.6\linewidth]{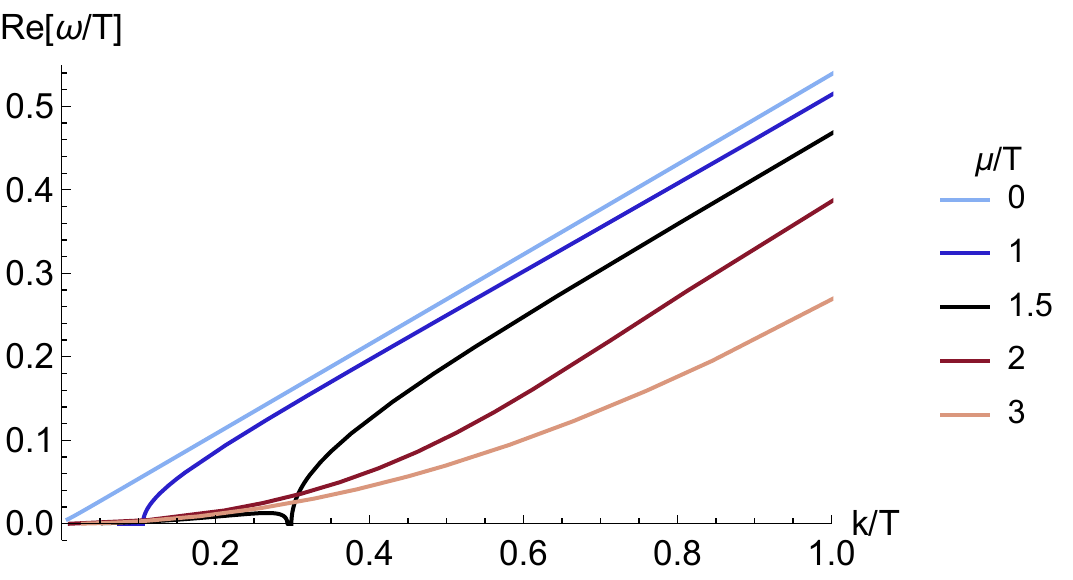}
\caption{\color{black} Zoom on the photon mode near the origin at large rigidity, $\alpha^3/T=3$. The mode transition from ordinary sound, to sound with a $k$-gap, to sound with a quadratic modulation at small $k$ when increasing $\mu/T$ is shown. In particular, an intermediary ``exotic'' dispersion is obtained for $\mu/T=1.5$. At large k, they all exhibit the same $v_{sound}$. \color{black}}
\label{fig:zoom2}
\end{figure}

\color{black} As a result, at large rigidity, this mode collision moves towards the origin and it creates the standard propagating sound mode $\omega \sim v_{sound} \,k$.
Further increasing the rigidity beyond this point produces a different strong modulation at small k, the quadratic dispersion seen on the right panel of figure~\ref{fig:phonons}. At small chemical potential, such a strong rigidity leads to non-trivial interactions between the phonon mode and the overdamped photon mode, resulting in a three-pole interaction similar to the one observed in~\cite{Gran:2018vdn}. At large $\mu/T$, the hydrodynamic mode becomes quadratic, $\mathrm{Re}(\omega)\sim k^2$, as discussed already in \cite{2017PhRvB..96p5115B}. This transition is shown in figure~\ref{fig:zoom2}.

\color{black}

\section{Conclusions}
In this work, motivated by the puzzling experimental results of \cite{Mitrano5392,husain2019crossover} and by the results of \cite{PhysRevLett.84.6026,doi:10.1063/1.5088141,PhysRevLett.97.115001}, we have analyzed in detail the spectrum of transverse collective modes in charged holographic systems taking Coulomb interactions into account. Our most interesting findings can be summarized as follows:
\begin{itemize}
    \item We observe a new mode repulsion mechanism in the presence of finite momentum relaxation, where the damped shear mode and the photon mode repel each other.
    \item We observe a propagation-to-diffusion crossover of the transverse collective modes at finite elasticity. This mechanism becomes more evident the stronger the electromagnetic interaction is.
    \item At large charge density, we observe a low frequency repulsion between the transverse collective mode and the photon mode. The spectrum of the modes has strong similarities with that of a charged Fermi liquid in the strongly interacting regime, $F_1^s>6$ (see figure~\ref{fig:Pines}), but with a crucial difference: the presence of an anomalous attenuation. The gapped photon mode is indeed damped also at zero momentum $k=0$ because of the interaction with a quantum critical continuum of states, as previously observed in holographic models for longitudinal collective modes \cite{Aronsson:2017dgf,Krikun:2018agd}. This feature might potentially be experimentally observed in strongly correlated materials in the quantum critical region, e.g.~in strange metals \cite{Mitrano5392,husain2019crossover}. \color{black} Recently, there have appeared concrete proposals for how to experimentally detect the transverse collective mode \cite{khoo2020quantum}, and it will be interesting to see if this leads to experimental results against which these holographic results can be compared.   \color{black}
\end{itemize}

From a theoretical point of view, there are several future directions which can be pursued with our methods. First, a more detailed comparison with the generalized global symmetry framework \cite{Hofman:2017vwr,Grozdanov:2017kyl} could be helpful in the direction of a unified theoretical understanding. Second, one could add a finite magnetic field to the framework including Coulomb interactions \cite{Aronsson:2017dgf,Aronsson:2018yhi,Mauri:2018pzq} to study magnetic screening. It should be straightforward to study spontaneous magnetization in holography and the emergence of magnons. We plan to consider some of these ideas in the near future.

\section*{Acknowledgements}
We thank Alberto Cortijo, Nabil Iqbal and Jan Zaanen for useful discussions about the topics discussed in this manuscript. M.B. acknowledges the support of the Spanish MINECO’s “Centro de Excelencia Severo Ochoa” Programme under grant SEV-2012-0249. 
U.G.~and M.T.~are supported by the Swedish Research Council.
\appendix
\section{Equations of motion and numerical methods}\label{app1}
We treat the system in linear response and consider the transverse sector, that is we add small perturbations to the background of section \ref{sec:model} of the form
\begin{align}
    g_{ty}\,\to\,g_{ty}(z)+\epsilon\,e^{-i\omega t+i k x}\,\delta\!g_{ty}(z)\,,\\
    g_{xy}\,\to\,g_{xy}(z)+\epsilon\,e^{-i\omega t+i k x}\,\delta\!g_{xy}(z)\,,\\
    A_{y}\,\to\,A_{y}(z)+\epsilon\,e^{-i\omega t+i k x}\,\delta\!A_{y}(z)\,,\\
    \Phi^{y}\,\to\,\Phi^{y}(z)+\epsilon\,e^{-i\omega t+i k x}\,\delta\!\Phi^{y}(z)\,,
\end{align}
where the equations of motion at order zero in $\epsilon$ are already solved by the background. At order one the equations of motion for the linear axions are presented in section \ref{sec:linax} and for the spontaneous symmetry breaking potential in section \ref{sec:ssbmodel}.

\subsection{Numerical methods}
The numerics were done in Mathematica, using the xAct \cite{xAct} and xTras \cite{Nutma:2013zea} packages.

The four coupled ordinary second order differential equations have four linearly independent solutions. One can be found analytically as a pure gauge mode. The remaining three are found numerically by imposing in-falling boundary conditions on the black hole horizon, and correspond to the values of $\delta\!A_y$, $\delta\!g_{xy}$ and $\delta\!\Phi^y$ on the horizon. To simplify the numerics, an expansion was made to a small offset from both the horizon and the boundary.

To find physical modes, we consider the full boundary value problem, with additional boundary conditions at the conformal boundary. There we impose Dirichlet boundary conditions on all fields except \color{black} $\delta\!A_y$\color{black}, on which we instead impose \eqref{eq:bcs}.
As the system is linear, requiring four boundary conditions on a linear combination of four independent solutions can be rephrased as evaluating the determinant of the corresponding matrix of boundary values, and find for what values of $(\omega,k)$ that the determinant,
\color{black}
\begin{equation}
\left|\begin{array}{cccc}
\delta\!g_{ty}(z)_{1} \;&\; \delta\!g_{xy}(z)_{1} \;&\; \delta\Phi^y(z)_1 \;&\; [\left(\omega^2-k^2\right)\delta\!A_y(z)+\lambda\,\delta\!A'_y(z)]_{1} \\
\delta\!g_{ty}(z)_{2} \;&\; \delta\!g_{xy}(z)_{2} \;&\; \delta\Phi^y(z)_2 \;&\; [\left(\omega^2-k^2\right)\delta\!A_y(z)+\lambda\,\delta\!A'_y(z)]_{2} \\
\delta\!g_{ty}(z)_{3} \;&\; \delta\!g_{xy}(z)_{3} \;&\; \delta\Phi^y(z)_3 \;&\; [\left(\omega^2-k^2\right)\delta\!A_y(z)+\lambda\,\delta\!A'_y(z)]_{3} \\
\delta\!g_{ty}(z)_{4} \;&\; \delta\!g_{xy}(z)_{4} \;&\; \delta\Phi^y(z)_4 \;&\; [\left(\omega^2-k^2\right)\delta\!A_y(z)+\lambda\,\delta\!A'_y(z)]_{4} 
\end{array}\right|_{z\to0}=0\,,
\end{equation}
\color{black}
is zero. Zero here being up to a numerically acceptable precision, which we chose to be significantly higher than necessary, simply to eliminate possible numerical artifacts.

\subsection{Linear axions}\label{sec:linax}
For the linear axions, $V(X)=X$, the equations of motion of the perturbations are:
\begin{flalign}
    \frac{\delta\!g_{ty} \left(-2 z f'+6 f-k^2 z^2+Q^2 z^4-6\right)}{z^2f}
   -\frac{k \omega  \delta\!g_{xy}}{f}-\frac{i \alpha  \omega  \delta\!\Phi^y}{f}\nonumber\\
  -2 Q z^2 \delta\!A_y'
   -\frac{2 \delta\!g_{ty}'}{z}
   +\delta\!g_{ty}''&=0\,,
\end{flalign}
\begin{flalign}
    \frac{k \omega  \delta\!g_{ty}}{f^2}
    +\frac{\delta\!g_{xy} \left(6 f^2-f \left(-z^2 f''+4 z f'+Q^2 z^4+\alpha ^2 z^2+6\right)+\omega ^2 z^2\right)}{f^2 z^2}
   +\frac{i \alpha  k \delta\!\Phi^y}{f}\nonumber\\
    +\left(\frac{f'}{f}-\frac{2}{z}\right) \delta\!g_{xy}'
   +\delta\!g_{xy}''&=0\,,
\end{flalign}
\begin{flalign}
    \frac{\delta\!A_y \left(\omega ^2-f k^2\right)}{f^2}
    +\frac{f' \delta\!A_y'}{f}
    -\frac{Q \delta\!g_{ty}'}{f}
    +\delta\!A_y''&=0\,,
\end{flalign}
and
\begin{flalign}
    \frac{\delta\!\Phi ^y \left(\omega ^2-f k^2\right)}{f^2}-\frac{i \alpha  \omega  \delta\!g_{ty}}{f^2}+\left(\frac{f'}{f}-\frac{2}{z}\right) {\delta\!\Phi ^y}'-\frac{i\alpha  k \delta\!g_{xy}}{f}+{\delta\!\Phi^y}''&=0\,.
\end{flalign}

\subsection{SSB model}\label{sec:ssbmodel}
In the spontaneous symmetry breaking case, $V(X)=X^3$, the equations of motion of the perturbations are similar, but altered slightly:
\begin{flalign}
-\frac{\delta\!g_{ty} \left(2 z f'-6 f+k^2 z^2-Q^2 z^4+8 \alpha ^6 z^6+6\right)}{f z^2}-\frac{k \omega  \delta\!g_{xy}}{f}-\frac{12 i \alpha ^5 \omega  z^4 \delta\!\Phi^y}{f}\nonumber\\
-2 Q z^2 \delta\!A_y'+\delta\!g_{ty}''-\frac{2 \delta\!g_{ty}'}{z}&=0\,,
\end{flalign}
\begin{flalign}
\frac{k \omega  \delta\!g_{ty}}{f^2}+\frac{\delta\!g_{xy} \left(6 f^2-f \left(-z^2 f''+4 z f'+Q^2 z^4+20 \alpha ^6 z^6+6\right)+\omega ^2 z^2\right)}{f^2 z^2}\nonumber\\
+\frac{12 i \alpha ^5 k z^4 \delta\!\Phi^y}{f}+\left(\frac{f'}{f}-\frac{2}{z}\right) \delta\!g_{xy}'+\delta\!g_{xy}''&=0\,,
\end{flalign}
\begin{flalign}
\frac{\delta\!A_y \left(\omega ^2-f k^2\right)}{f^2}+\frac{f' \delta\!A_y'}{f}-\frac{Q \delta\!g_{ty}'}{f}+\delta\!A_y''&=0\,,
\end{flalign}
and
\begin{flalign}
\frac{\delta\!\Phi ^y \left(\omega ^2-f k^2\right)}{f^2}-\frac{i \alpha  \omega  \delta\!g_{ty}}{f^2}+\left(\frac{f'}{f}+\frac{2}{z}\right) {\delta\!\Phi ^y}'-\frac{i \alpha  k \delta\!g_{xy}}{f}+{\delta\!\Phi ^y}''&=0\,.
\end{flalign}

\bibliographystyle{JHEP}
\bibliography{plas2}
\end{document}